\documentclass[11pt]{article}
\usepackage[dvips]{graphicx}
\usepackage{a4wide,cite,float,latexsym}
\usepackage[french,english]{babel}
\usepackage{amsmath}
\usepackage{epsfig}
\newcommand{\lagr}{{\cal L}}
\newcommand{\Veff}{{\cal V}}    
\newcommand{\Aeff}{{\cal A}}
\newcommand{\calo}{{\cal O}}
\newcommand{\Frac}[2]{\frac{\displaystyle #1}{\displaystyle #2}}

\begin{document}

\begin{titlepage}

\begin{center}
{\LARGE\bf \boldmath{$K^L_{\mu3}$} decay: A Stringent Test of Right-Handed Quark Currents}\\[12mm]

{\normalsize\bf V\'eronique Bernard~${}^{a,}$\footnote {Email:~bernard@lpt6.u-strasbg.fr},
Micaela Oertel~${}^{b,}$\footnote {Email:~oertel@ipnl.in2p3.fr},
Emilie Passemar~${}^{c,}$\footnote{Email:~passemar@ipno.in2p3.fr} and Jan Stern~${}^{c,}$\footnote{Email:~stern@ipno.in2p3.fr}} \\[4mm]

{\small\sl ${}^{a}$ Universit\'{e} Louis Pasteur, Laboratoire de Physique Th\'{e}orique,} \\
{\small\sl 3-5 rue de l'Universit\'{e}, F-67084 Strasbourg, France}\\
{\small\sl ${}^{b}$ LUTH, Observatoire de Paris, F-92195 Meudon, France} \\
{\small\sl ${}^{c}$ Groupe de Physique Th\'{e}orique, IPN,
           Universit\'{e} de Paris Sud-XI, F-91406 Orsay, France}\\[12mm]
\end{center}

\noindent{\bf Abstract:}
\noindent Clean tests of a small admixture of right-handed quark currents directly
coupled to the standard $W$ are still lacking. We show that such non-standard
couplings can be signifi-cantly constrained measuring the value of the scalar 
$K\pi$ form factor at the Callan-Treiman point to a few percent.
A realistic prospect of such a measurement in $K^L_{\mu3}$ decay
based on an accurate dispersive representation 
of the scalar form factor is presented. The inadequacy of the currently
used linear parametrisation is explained and illustrated using recent
KTeV data. We briefly comment on the charged Kaon mode.  

\end{titlepage}
\pagenumbering{arabic}
\renewcommand{\thefootnote}{\arabic{footnote}}
\parskip12pt plus 1pt minus 1pt
\topsep0pt plus 1pt
\setcounter{totalnumber}{12}

{\bf I.}       In this letter we propose a new dedicated test of       
(the absence of)
charged right-handed currents (RHCs) involving light quarks. We show
how high statistics measurements~\cite{Alexopoulos:2004sy,NA48} of Dalitz
distributions in the $K^L_{\mu 3}$ decay, $K_L \to \pi^{\pm} \mu^{\mp}\nu$,
can be used to extract the value $\mathrm{C} = f(\Delta_{K\pi})$ of the
scalar $K \pi$ form
factor at the Callan-Treiman point $\Delta_{K\pi} = m_{K^0}^2 -m_{\pi^+}^2$
and how this information
constrains the effect of
$\bar{u}d$ and $\bar{u}s$ RHCs. The method is model independent. It is based
on the observation that standard dispersive technics and the known low-energy
$K\pi$ phases lead to an accurate parametrization of the scalar $K\pi$
form factor in terms of a single parameter $\mathrm{C}$ subject to experimental
determination. The related theoretical uncertainties are under control
and they remain small compared with the possible signal of RHCs. We will 
comment on charged K-decays shortly in section {\bf VI}.

{\bf II.}           In the past, tests of RHCs have often been considered
in connection
with left-right symmetric extensions of the Standard Model
(SM)~\cite{Mohapatra:1974gc}, yielding a lower bound
on the mass $M_{W_R}$ of the hypothetical vector boson mediating the
RH weak interactions~\cite{Mohapatra:1983aa}. Different models leading to 
RHCs through mixing with heavy exotic fermions~\cite{delAguila:2000rc} have been considered, too. 
Independently of any specific models, RHCs
naturally arise in Low Energy Effective Theories which below some
scale $\Lambda_W$ operate with SM degrees of freedom and
symmetries. Such RHC interactions are not necessarily mediated by an extra gauge
boson $W_{R\mu}$ and they need not be concerned by phenomenological lower bounds on $M_{W_R}$.
Indeed, there exists a unique $\mathrm{SU}(2)\times \mathrm{U(1)}$
gauge-invariant operator~\cite{BW86}
\begin{equation} 
O_{\mathrm{RHC}} = \frac{1}{\Lambda^2} (\bar{U}_R\gamma^\mu D_R) \phi_r \epsilon^{rs} (D_\mu \phi)_s
\label{RHCop}
\end{equation}
describing a direct coupling of $W^\mu$ with the RHC
$\bar{U}_R\gamma_\mu D_R$. The presence of the Higgs doublet $\phi_r
(r = 1,2)$ in Eq.~(\ref{RHCop}) suggests that the actual strength of
this operator depends on the mechanism of electroweak symmetry
breaking.\newline
\indent If the symmetry is linearly realized and the light Higgs particle
exists, the operator Eq.~(\ref{RHCop}) is just one among the 80
independent gauge invariant operators of the mass (or UV) dimension
$D=6$~\cite{BW86} and it is not easy to disentangle their respective
dimensional suppression.\newline
\indent In the opposite case of a Higgs-less effective theory~\cite{HS1,Csaki:2003sh}, only the
three Goldstone bosons contained in the complex Higgs doublet $\phi_r$
remain in Eq.~(\ref{RHCop}) and the operator $O_\mathrm{RHC}$ becomes
of the chiral (infrared) dimension $d = 2$, i.e., it is not
dimensionally suppressed anymore. Instead, its suppression is now
related to its symmetry properties with respect to the higher
non-linearly realized gauge symmetry characteristic of Higgsless
vertices of the SM~\cite{HS1}. As a result the RHC interaction now
appears already at the NLO: it represents a genuine effect beyond the
SM that is potentially more important than the loop
corrections\footnote{Notice that in the SM 1PI vertices bewteen the
$W$ and right-handed fermions ${i,j}$ are induced at one-loop order
and are proportional to $m_i m_j$. This is the reason why we are
particularly interested in RHC of light quarks.}. In the framework of a 
systematic low-energy expansion, one should first consider observables
that are {\it linear} in the operator Eq.~(\ref{RHCop}) (or in its Higgs-less
analogue given in~\cite{HS1}). \newline
\indent There exists a compelling experimental evidence against charged
RHCs in the {\it lepton sector} based on polarisation measurements in $\mu$-decay, $\tau$-decay
and $\beta$-decay~\cite{Mohapatra:1983aa,Beg:1977ti}. Recent findings on neutrino 
mixings and masses however suggest that quark and lepton sectors and, in particular, the
corresponding RHCs interactions need not be alike. If the right (Majorana)
neutrino is heavy compared to the scale $\Lambda_W$ of the Effective theory,
there will be no RHCs visible at low energy. In the opposite case, the lepton
sector should enjoy an extra symmetry (not present for quarks) that suppresses
the neutrino Dirac mass and thereby the leptonic charged RHCs as well~\cite{HS1}.
(The simplest example of such a discrete symmetry is the $\nu_R$ sign flip
symmetry $\nu_R \to -\nu_R$ introduced in~\cite{HS1}.) Notice that the disymmetry
between quark and lepton couplings beyond the leading order would generate
an anomaly which in the Effective theory is compensated by the Wess-Zumino
term constructed in~\cite{HS2}. In particular, there does not seem to be any
obvious consistency or plausibility argument against the quark charged
RHCs Eq.~(\ref{RHCop}) even if the latter is absent for leptons. \newline
\indent Surprisingly enough the available experimental constraints on first order
RHC effects of quarks remain so far rather meagre. They suffer either from the lack
of precision or from the excess of model dependence facing
non-perturbative QCD effects. Global fits to electroweak precision
data based on electroweak effective
Lagrangians~\cite{Burgess:1993vc} usually do not strongly
constrain the operator Eq.~(\ref{RHCop}). Some time ago, the CDHS
collaboration has reported a dedicated test of RHCs, later confirmed by CCFR~\cite{Abramowicz:1981iq}, based
on the $\mathrm{y}$-dependence of $\nu (\bar{\nu})$ DIS off valence
quarks. Unfortunately, only the square of the RHC operator,
Eq.~(\ref{RHCop}), contributes to the leading twist. Such contribution of RHCs
is strongly suppressed and it can be hardly disentangled from higher twist
left-handed contributions. It has been further observed that a RHC interaction
would alter the chiral structure of the tree-level effective weak
Hamiltonian and the corresponding consequences of soft-pion theorems
for weak $K$ and hyperon decays~\cite{Donoghue:1982mx} in
an experimentally relevant way. However, the upper bounds on RHCs
derived from this observation did not consider long distance chiral
loop corrections (such as final state interaction) which are known to
be rather important and can easily upset small tree-level effects. 
The strongest constraint on left-right symmetric models is known to arise 
from the $W_{R\mu}$-exchange contribution to the $K^0-\Bar{K}^0$ mixing~\cite{Beall:1981ze}.
However, in an effective theory where a (light) $W_{R\mu}$ is absent, the leading RHC effect
arises from the $\bar{s}d \leftrightarrow \bar{d}s$ box diagram with an
insertion of a {\it single RHC vertex} Eq.~(\ref{RHCop}). Due to Lorentz invariance,
the resulting four-fermion operator in $H_W^{\Delta S=2}$ necessarily involves
derivatives and it is suppressed by the external momentum scale in addition to 
the suppression by the small coefficient of the operator in Eq.~(\ref{RHCop}).
A similar argument applies to $\Delta S=1$ FCNC such as $K\rightarrow \pi \nu \bar{\nu}$.
All such contributions of RHCs to FCNC processes are expected to be smaller
than in the SM.

                It is convenient to write the effective CC interaction vertex
in a matrix notation
\begin{equation}
\lagr_{CC} = \tilde{g}[l_\mu +\frac{1}{2}\bar{\mathrm{U}}( \Veff_{eff} \gamma_{\mu}  +
\Aeff_{eff}\gamma_{\mu} \gamma_5 ) \mathrm{D} ] W^\mu  +  h.c.\
,\,
\mathrm{U}=
\begin{pmatrix}
u\\ c \\t
\end{pmatrix},\,
\mathrm{D}=
\begin{pmatrix}
d\\ s \\b 
\end{pmatrix},
\label{lagrCC}
\end{equation}
and $\Veff_{eff}$, $\Aeff_{eff}$ are complex $3 \times 3$ effective      
coupling matrices. In the Standard Model one has
\begin{equation}
\Veff_{eff} = - \Aeff_{eff} = V_{CKM},
\label{SM}
\end{equation}
where $V_{CKM}$ is the unitary flavour mixing matrix. Recently, much
effort has been devoted to experimentally test the unitarity of
$V_{CKM}$. The issue of such tests usually depends on the stage of our
theoretical knowledge of corresponding hadronic matrix elements. For
instance, in the case of light quark elements $V^{ud}$ and $V^{us}$
the CKM unitarity is not yet established: whereas lattice calculations
of the $K_{e3}$ decay form factor $f_{+}(0)$~\cite{Becirevic:2004ya}
are compatible with the first row of CKM unitarity, the two-loop $\chi
PT$ calculation~\cite{Bijnens:2003uy} indicates a possible violation
of the latter by as much as
$2.6~\sigma$~\cite{Cirigliano:2005xn}. Furthermore, there is
practically no significant test of the relation $\Veff_{eff} = -
\Aeff_{eff}$, i.e. of the absence of RHCs.  There is an intermediate
step between the SM case Eq.~($\ref{SM}$) and the completely general
effective couplings Eq.~($\ref{lagrCC}$). The departure from the SM
can still be universal up to and including the NLO, meaning that there
exists a chiral flavour basis in which both $\Veff_{eff}$ and
$\Aeff_{eff}$ are proportional to the unit matrix. (In other words,
all flavor symmetry breaking can be transformed from vertices to the
mass matrix.) This property is shared by many models with minimal
flavor violation~\cite{D'Ambrosio:2002ex}. It can be equivalently expressed in terms of
effective couplings as ($i= u , c ,t; j= d ,s ,b $)
\begin{equation}
\Veff_{eff}^{ij}=(1+\delta) V_L^{ij}+\epsilon V_R^{ij}+\mathrm{NNLO}\,,
\hspace{1cm}
\Aeff_{eff}^{ij}=-(1+\delta) V_L^{ij}+\epsilon V_R^{ij}+\mathrm{NNLO}.
\label{effective couplings}
\end{equation}
Here, $\delta$ and $\epsilon$ are two small parameters measuring the departure
from the SM , whereas $V_L$ and $V_R$ are two a priori independent unitary
matrices arising from the diagonalisation of the generic quark mass matrix.
As already discussed, we do not introduce charged leptonic RHCs: in Eq.~($\ref{lagrCC}$)
$l_{\mu}$ stands for the standardly normalized V-A lepton current. A few
direct first order constraints on the parameters $\epsilon$ and $\delta$ at the percent 
level are conceivable. They can be obtained from selected tree-level
semi-leptonic processes in which QCD effects are under theoretical control.
This is true, in particular, for inclusive hadronic $\tau$-decays  whose impact 
on the parameters $\epsilon$ and $\delta$ will be discussed separately~\cite{BOPS06}.
Here, we concentrate on the $K_{\mu3}$ decay which could provide one of the most stringent
probes of RHCs. To the best of our knowledge, none of these tests has been considered
previously.   

  {\bf  III.} We consider the hadronic matrix element describing       
the $K^0_{\mu3}$ decay:
\begin{equation}
\langle \pi^-(p') | \bar{s}\gamma_{\mu}u | K^0(p)\rangle = 
(p'+p)_\mu\  f^{K^0\pi^-}_+ (t) + (p-p')_\mu\  f_-^{K^0\pi^-} (t),         
\label{hadronic element}
\end{equation}
where $t=(p'-p)^2=(p_\mu+p_\nu)^2$. 
 The vector form factor $f^{K^0\pi^-}_+ (t)$ represents
the P-wave projection of the crossed channel matrix element
$\langle 0 |\bar{s}\gamma_{\mu}u | K\pi \rangle$, whereas                
the S-wave projection is described by the scalar form factor:
\begin{equation}
f_S (t) = f^{K^0\pi^-}_+ (t) + \frac{t}{m^2_{K^{0}} - m^2_{\pi^{-}}} f^{K^0\pi^-}_-(t).
\label{defffactor}
\end{equation}
In the experimental study of decay distributions one usually factorizes from
both form factors $f^{K^0\pi^-}_+(t)$ and  $f^{K^0\pi^-}_S(t)$ the common
factor $f^{K^0\pi^-}_+(0)$ to normalize them to 1 at $t=0$. We thus concentrate
on the normalized scalar form factor    
\begin{equation}
f(t)=\frac{f^{K^0\pi^-}_S(t)}{f^{K^0\pi^-}_+(0)}\ \ ,\ \ f(0)= 1 .
\label{defnffactor}
\end{equation}
The Callan-Treiman low-energy theorem (CT)~\cite{Dashen:1969bh} fixes the value of
$f(t)$ at the point $t=\Delta_{K\pi}=m_{K^{0}}^2-m_{\pi^+}^2$ in the 
$\mathrm{SU}(2)\times \mathrm{SU}(2)$ chiral limit. We can write
\begin{equation}
\mathrm{C}=f(\Delta_{K\pi})=\frac{F_{K^+}}{F_{\pi^+}}\frac{1}{f_{+}^{K^0\pi^-}(0)}+  
\Delta_{CT},
\label{C}
\end{equation}
where the CT discrepancy $\Delta_{CT}$ defined by Eq.~($\ref{C}$) is
expected to be small and eventually calculable in $\chi PT$.
It is proportional to $m_u$ and/or $m_d$. In the
limit $m_d=m_u$ at the NLO in $\chi PT$ one has for the CT discrepancy
$\Delta_{CT}^{\mathrm{NLO}}= - 3.5 \times 10^{-3}$~\cite{Gasser:1984ux}. We will come back
to it in section {\bf IV}.\\ 
As a next step, we express $\mathrm{C}=f(\Delta_{K\pi})$ in terms of
measured branching ratios and of the CC effective couplings defined in
Eq.~($\ref{effective couplings}$).  From the branching ratio 
Br~$\Frac{K^+_{l2}(\gamma)}{\pi^+_{l2}(\gamma)}$~\cite{Marciano:2004uf},
one gets
\begin{equation}
\Bigl{|}{\frac{F_{K^+} \Aeff_{eff}^{us}}{F_{\pi^+} \Aeff_{eff}^{ud}}}\Bigr{|}^2 = 0.07602(23)(27),
\label{BR}
\end{equation}
whereas the weighted average of the 3 compatible most recent
measurements of the inclusive decay rate $K^L_{e3} (\gamma)$ by
KTeV~\cite{Alexopoulos:2004sw}, NA48~\cite{Lai:2004bt}, and
KLOE~\cite{Ambrosino:2005ec} leads to
\begin{equation}
|f_+^{K^0\pi^-}(0) \Veff_{eff}^{us}| = 0.21619(55).
\label{f+Vus}
\end{equation}
The expression for C can now be rewritten as:
\begin{equation}
\mathrm{C}= B_{exp}~r + \Delta_{CT}~.
\label{Cbis}
\end{equation}
The branching ratio  $B_{exp}=\Bigl{|}\frac{F_{K^+} \Aeff_{eff}^{us}}
{F_{\pi^+} \Aeff_{eff}^{ud}}\Bigr{|}
\frac{1}{|f_+^{K^0\pi^-}(0) \Veff_{eff}^{us}|}|\Veff_{eff}^{ud}|$, where 
$|\Veff_{eff}^{ud}|$ is precisely known from superallowed
$0^+\to 0^+$ nuclear $\beta$-decays~\cite{Hardy:2004id} with the
recently updated accuracy~\cite{Marciano:2005ec}
\begin{equation}
|\Veff_{eff}^{ud}| = 0.97377(26)~,
\label{Vud}
\end{equation}
is $B_{exp}=1.2418 \pm 0.0043$ using Eqs.~($\ref{BR}$), 
($\ref{f+Vus}$) and~($\ref{Vud}$).
The parameter $r$ is given by the RHCs effective couplings
\begin{equation}
r = \Bigl{|}\frac{\Aeff_{eff}^{ud} \Veff_{eff}^{us}}{\Veff_{eff}^{ud}
\Aeff_{eff}^{us}}\Bigr{|} 
= 1 + 2 (\epsilon_{S}-\epsilon_{NS})+ \calo(\epsilon^2)~,
\label{r}
\end{equation}
where 
\begin{equation}
 \epsilon_{NS}= \epsilon\ \mathrm{Re} \Bigl{(}\frac{V_R^{ud}}{V_L^{ud}}\Bigr{)},\ \
 \epsilon_{S} = \epsilon\ \mathrm{Re} \Bigl{(}\frac{V_R^{us}}{V_L^{us}}\Bigr{)}
 \label{epsilon}
\end{equation}
represent the strengths of $\bar ud$ and $\bar us$ RHCs, respectively.
The deviation of $r$ from its SM value $r_{SM}= 1$ signalizes the
presence of RHCs, i.e., a non-vanishing parameter $\epsilon$. Notice however,
that the inverse is not true: RHCs characterized by the mixing matrix
$V_R$ aligned with the CKM left handed matrix, i.e.
$ V_R = \exp (i \omega) V_L$~,
would not show up in the parameter $r$ and would escape the detection
in $K^L_{\mu3}$ decays.\\
Using the experimental number for the 
branching ratio mentionned above one gets:
\begin{equation} 
\mathrm{ln C} = 0.2166 \pm 0.0035 + 
\tilde\Delta_{CT} + 2 (\epsilon_S - \epsilon_{NS})+\calo(\epsilon^2)~, 
\label{lnC}
\end{equation} 
where $\tilde\Delta_{CT}= \Delta_{CT}/B_{exp}$.
We now ask how big the effect of RHCs should be to be seen measuring
$\mathrm{ln C}$ and taking into account the experimental and
theoretical ($\tilde\Delta_{CT}$) uncertainties quoted in Eq.~($\ref{lnC}$).
We take as an example the Effective Higgs-less
theory, where RHCs should appear at the NLO before loop
effects~\cite{HS1}. There, the order of magnitude estimate based
on the momentum and spurion power counting~\cite{HS1} suggests
$\epsilon \sim 0.005 \div 0.010$. We have performed a separate
analysis of the effective CC couplings in hadronic $\tau$-decays~\cite{BOPS06}
leading to results compatible with a similar
range of values for $\epsilon_{NS}$ but giving no precise information
on $\epsilon_S$.  Taking the extreme possibility $\epsilon_S =
-\epsilon_{NS}$, one can foresee the effect of RHCs in $\mathrm{ln C}$
as large as $0.02 \div 0.04$. (As already pointed out, $\epsilon_S =
\epsilon_{NS}$ would imply no effect even if RHCs were actually
present.) We conclude that an effect of RHCs significantly larger than
the uncertainties in $\mathrm{ln C}$ cannot be a priori
excluded. Hence, a measurement of $\mathrm{ln C}$ to $5 \div 10\%$
could represent a relevant experimental information/bound on RHCs
interactions.
 
\begin{figure}[h]
\vspace{0.1cm}
\begin{center}
\includegraphics*[scale=1.]{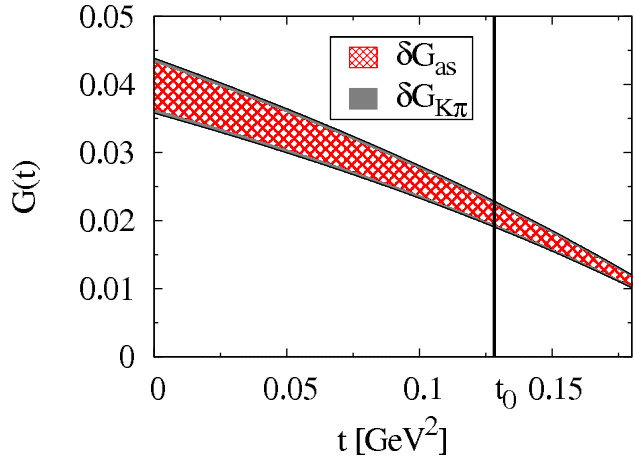}
\caption{G(t) with the uncertainties $\delta \mathrm{G}_{as}$ 
and $\delta \mathrm{G}_{K \pi}$ added in quadrature.}
\label{figureg}
\end{center}
\end{figure}

{\bf IV. }  In the sequel we propose a new (exact) parametrization of the
 scalar form factor $f(t)$ which should allow a model independent extraction
 of $\mathrm{ln C}$ from experimental data. We write a dispersion relation for
 $\ln f(t)$ subtracted at the points $t = 0$ and $t = \Delta_{K\pi}$.
 One usually assumes that $f(t)$ does not have zeros. In this case, one can write:
\begin{equation}
f(t)=\exp\Bigl{[}\frac{t}{\Delta_{K\pi}}(\mathrm{lnC}-\mathrm{G(t)})\Bigr{]}~,
\label{Dispf}
\end{equation}
where
\begin{equation} 
\mathrm{G(t)}=\frac{\Delta_{K\pi}(\Delta_{K\pi}-t)}{\pi}\int_{t_{\pi K}}^{\infty}
\frac{dx}{x}\frac{\phi(x)}
{(x-\Delta_{K\pi})(x-t-i\epsilon)}\\~,
\label{G}
\end{equation} 
$t_{\pi K}$ is the threshold of $\pi K$ scattering and $\phi(t)$ is the
phase of $f(t)$:
\begin{equation}
f(t)=|f(t)| \exp(i\phi(t))~.
\end{equation}
As $t \to - \infty$, one expects $f(t) =
\calo(1/t)$~\cite{Lepage:1979zb}. Consequently, for large $t$, the phase $\phi(t) \to
\pi$, implying a rapid convergence of the twice subtracted dispersion
integral, Eq.~(\ref{G}). According to Watson's theorem, the phase $\phi(t)$
should coincide with the S-wave $\mathrm{I} = 1/2$ $K\pi$ scattering
phase $\delta_{K\pi}(t)$ for sufficiently low energies.  As observed
experimentally~\cite{Estabrooks:1977xe}, the $\mathrm{I} = 1/2$ S-wave $K\pi$ scattering
amplitude is to a very good approximation elastic up to the
c.m. energy $\mathrm{E}=1.67$ GeV, where the phase almost reaches
$\pi$. After this point the phase drops out and the inelasticity sets
in.  In the following, we will assume that up to this energy, i.e., for
$t < \Lambda = 2.77\ \mathrm{GeV}^2$ one has $\phi(t) =
\delta_{K\pi}(t)$. Above $\Lambda$, we take $\phi(t) = \pi$ and we
include the possible deviation from this asymptotic estimate into the
error. As a result, $\mathrm{G(t)}$ can be decomposed as
\begin{equation}
\mathrm{G(t)}=\mathrm{G}_{K\pi}(\Lambda,t) + \mathrm{G}_{as}(\Lambda,t) \pm \delta \mathrm{G}(t).
\label{G decomposed}
\end{equation}
The first term represents the integral Eq.~($\ref{G}$) from $t_{\pi
K}$ up to $\Lambda$ with $\phi(t)$ replaced by the scattering phase
$\delta_{K\pi}$. The latter is precisely known in the whole
integration range down to the threshold from matching the solution of
Roy-Steiner Equations with $K\pi \to K\pi,~\pi\pi \to \bar KK$ and
$\pi\pi \to \pi\pi$ scattering data available at higher energy.
We refer the reader to the work~\cite{Buettiker:2003pp} containing
all the details of the Roy-Steiner analysis of the $\pi K$ scattering
amplitude and the resulting phase $\delta_{K\pi}$ we use in evaluating
$\mathrm{G}_{K\pi}(\Lambda,t)$. The second term $\mathrm{G}_{as}(\Lambda,t)$ on
the RHS of Eq.~($\ref{G decomposed}$) stems from the asymptotic tail
of the dispersive integral Eq.~($\ref{G}$) between $\Lambda$ and
$\infty$ assuming that in this range the phase of the form factor can
be replaced by its asymptotic value $\pi$. Explicitly,
\begin{equation}
\mathrm{G_{as}}(\Lambda,t) = \frac{\Delta_{K \pi}}{t}\mathrm{ln}\Bigl{(}1-\frac{t}{\Lambda}\Bigr{)}-
\mathrm{ln}\Bigl{(}1-\frac{\Delta_{\pi K}}{\Lambda}\Bigr{)}~.
\label{Gas}
\end{equation}
One can easily check that for $\Lambda = 2.77\ \mathrm{GeV}^2$ and in
the relevant range $0 < t < \Delta_{K\pi}$ the asymptotic contribution
to G is tiny $0 < \mathrm{G}_{as}(\Lambda,t) < 0.0036$, compared with
$\mathrm{ln C} \sim 0.20$.  Finally, there are two distinct sources of
uncertainty $\delta \mathrm{G}$ in Eq.~($\ref{G decomposed}$). The
first, $\delta \mathrm{G}_{K\pi}$, arises from the error on the low-energy
$K\pi$ phase shifts entering the dispersive integral
$\mathrm{G}_{K\pi}$. $\delta \mathrm{G}_{K\pi}$ is estimated
inspecting the propagation of errors in the experimental input into
the solution of Roy-Steiner equations. and varying the corresponding
matching point\footnote{We are indebted to Bachir Moussallam for the
help with this error analysis following Ref.~\cite{Buettiker:2003pp}}.
In this way $\delta \mathrm{G}_{K\pi}(t)$ can be obtained point by
point together with the corresponding correlation matrix. Here we just
mention the uniform bound
\begin{equation}
\delta \mathrm{G}_{K\pi}(t) \le 0.05 \times  \mathrm{G}_{K\pi}(t)~,
\label{dGKpi}
\end{equation}
which faithfully resumes the effect of the uncertainty. The second source
of error stems from the unknown high energy phase $\phi(t)$ of the form
factor. For $t > \Lambda = 2.77\ \mathrm{GeV}^2$ we propose a generous estimate
$\phi(t) = \pi \pm \pi$  which amounts to 
\begin{equation}
\delta \mathrm{G}_{as} (t) = \mathrm{G}_{as} (t)~,
\label{dGas}
\end{equation}
where $\mathrm{G}_{as} (t)$ is given in Eq.~($\ref{Gas}$).
The resulting function $\mathrm{G(t)}$ is shown in Fig.~\ref{figureg}
together with the two uncertainties added quadratically.
One observes
that in the whole physical region of the $K^L_{\mu3}$ decay the
function $\mathrm{G(t)}$ does not exceed $20\%$ of the expected value
of $\mathrm{ln C}$. The uncertainty $\delta \mathrm{G_{as}}$ which
clearly dominates could be further reduced using a somewhat model
dependent multi channel Omnes-Mushkelishvili construction extending 
the description of the scalar form factor to higher energies. Such a construction
has been presented in Ref.~\cite{Jamin:2001zq} and in principle, it
allows to infer the phase of the form factor above the elastic
region. We have used this phase to check, that in the low-energy
region of interest the model of Ref.~\cite{Jamin:2001zq} reproduces
our function $\mathrm{G(t)}$ within errors.\newline
\indent For practical purposes we give a simple parametrization of
$\mathrm{G(t)}$ in the physical region
$ m_{\mu}^2 < t < t_0 = (m_{K^0} - m_{\pi^+} )^2$. Denoting $x = t/t_0$, the
true function $\mathrm{G(t)}$ is to a very good accuracy reproduced by
\begin{equation}
\mathrm{G_P(t)} = x D + (1 - x) d  + x (1 - x) k~,
\label{GP}
\end{equation}
where $d = \mathrm{G}(0)$, $D =\mathrm{G(t_0)}$ and $k$ is obtained
from the constraint $\mathrm{G}(\Delta_{K\pi}) = 0$. The central
values of the three parameters $d$, $D$ and $k$ are collected in Table
1 together with the corresponding errors arising from $\delta
\mathrm{G}_{K\pi}$ and from $\delta \mathrm{G_{as}}$.
The uncertainties shown in Table~1 are correlated as implied by
Eqs.~(\ref{dGKpi}) and (\ref{dGas}). For the central values of Table~1, the
deviation of the polynomial approximation $\mathrm{G_P(t)}$ from
the exact function G(t) does not exceed $1\%$ of G(t) in the whole physical
region.
\begin{table}[h!]
\begin{center}
\begin{tabular}{l|c|c|c|c}
 & Central value & $\delta \mathrm{G}_{as}$ & $\delta \mathrm{G}_{K\pi}$ \\ 
\hline 
$d$& 0.0398 &0.0036 & 0.002\\ 
\hline
$D$ &0.0209 &0.0016 & 0.001 \\ 
\hline
$k$ &0.0045 &0.0001 &\\ 
\hline
\end{tabular}
\end{center}
\caption {Coefficients arising in the parametrization $\mathrm{G_P}$ with their uncertainties.}
\end{table}

It should be stressed that an accurate determination of $\mathrm{lnC}$
using the dispersive representation Eq.~($\ref{Dispf}$) in the fit of
measured distributions only allows to infer from Eq.~($\ref{lnC}$) the
combination 
\begin{equation} 
\Delta \epsilon = 2 (\epsilon_S - \epsilon_{NS}) + \tilde \Delta_{CT}
\label{deps} 
\end{equation} 
\noindent of the Callan Treiman discrepancy $\tilde{\Delta}_{CT}$ as defined by
Eq.~($\ref{C}$) ($\tilde{\Delta}_{CT} = \Delta_{CT}/B_{exp}$) and the RHCs parameter
$\epsilon_S - \epsilon_{NS}$. In order to isolate the latter, the theoretical
input of the former is required. Including isospin breaking effects at the order 
$\calo(p^4,(m_d-m_u)p^2,e^2p^2)$ and varying the input parameters ($R=\frac{m_s-\hat{m}}{m_d-m_u}$,
$F_{\pi}$, $L_9$ and $L_5$) one obtains from the $\chi PT$ formula displayed in~\cite{Cirigliano:2001mk}
$\Delta_{CT}$ in the range $-0.005 \div -0.001$ to be compared with the Gasser-Leutwyler estimate~\cite{Gasser:1984ux}
given previously. This numerical result is somewhat sensitive to the use of 
the Gell-Mann-Okubo formula for the $\eta$ mass at that order,
indicating a possible im-

\begin{figure}[h!]
\vspace{0.1cm}
\begin{center}
\includegraphics*[scale=0.9]{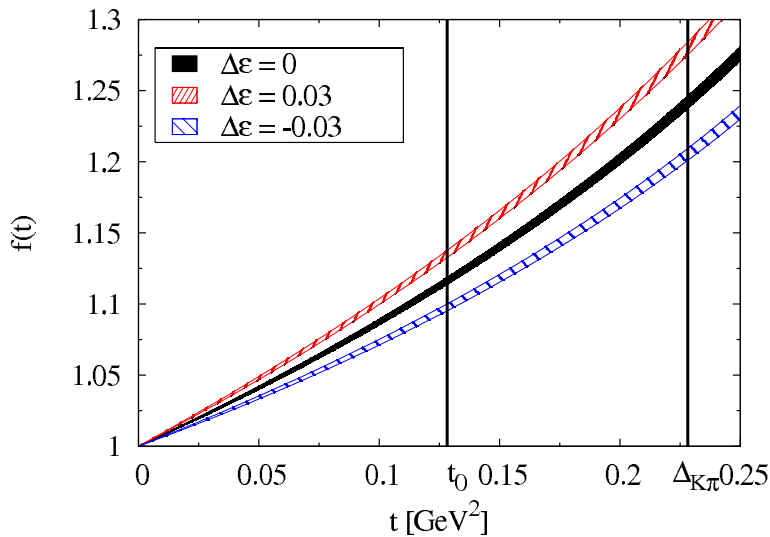}
\caption{The scalar form factor $f(t)$. Central curve: $\Delta \epsilon = 0$,
top curve: $\Delta \epsilon = 0.03$,
bottom curve: $\Delta \epsilon = -0.03$ with uncertainties from experimental branching ratios and
from G(t) added in quadrature.}  
\label{figuref}
\end{center}
\end{figure}                       

\noindent portance of the $\calo(p^6)$ contribution. The precision of this
 estimate could be improved once the two loop $\chi PT$ analysis of $K_{l3}$
form factors~\cite{Bijnens:2003uy} will be completed determining the relevant $\calo(p^6)$
LECs. At present our result suggests $\tilde{\Delta}_{CT}$ 
an order of magnitude smaller than the possible values of $\Delta \epsilon$ indicated by recent data~\cite{Alexopoulos:2004sy} (see section {\bf V}).

In Fig.~\ref{figuref} we show the sensitivity of the normalized scalar form factor $f(t)$
to the parameter $\Delta \epsilon$, Eq.~($\ref{deps}$).
The central curve shows the form factor $f(t)$ for $\Delta \epsilon = 0$ as given by the
dispersive representation Eq.~($\ref{Dispf}$) with the use of Eq.~($\ref{lnC}$).
The error arising from experimental branching ratios as well as the error $\delta
\mathrm{G(t)}$ are superposed to the curve.
The two extreme curves with their error bars represent the form factor in the cases
$\Delta \epsilon = \pm 0.03$. Once more it is seen that the
uncertainties are much smaller than the possible signal of RHCs.\\
\indent We add a comment on the parametrization of the form factor $\hat{f}_{+}(t)$. 
At present, it seems difficult to construct a one-parametrical representation
of $\hat{f}_{+}(t)$ which would be as accurate as the representation Eq.~($\ref{Dispf}$)
of the scalar form factor $f$. The reason is that the inelasticity in the $K \pi$-scattering
P-wave sets in at lower energies and furthermore, the experimental information on 
$K^{+}\pi^{+} \to K^{+}\pi^{+}$ P-wave is still missing. Under these circumstances, it 
seems preferable to continue using the simple two-parametrical representation~\cite{Alexopoulos:2004sy}:
\begin{equation}
\hat{f}_{+}(t)=\frac{f_{+}^{K^0\pi^{-}}(t)}{f_{+}^{K^0\pi^{-}}(0)}=
1+\lambda_{+}\frac{t}{m_{\pi}^2}+\frac{1}{2}\lambda'_{+}\Bigl{(}\frac{t}{m_{\pi}^2}\Bigr{)}^2~,
\label{f+}
\end{equation}
keeping in mind the probable relevance of the curvature term.

\begin{figure}[h!]
\vspace{0.1cm}
\begin{center}
\includegraphics*[scale=0.9]{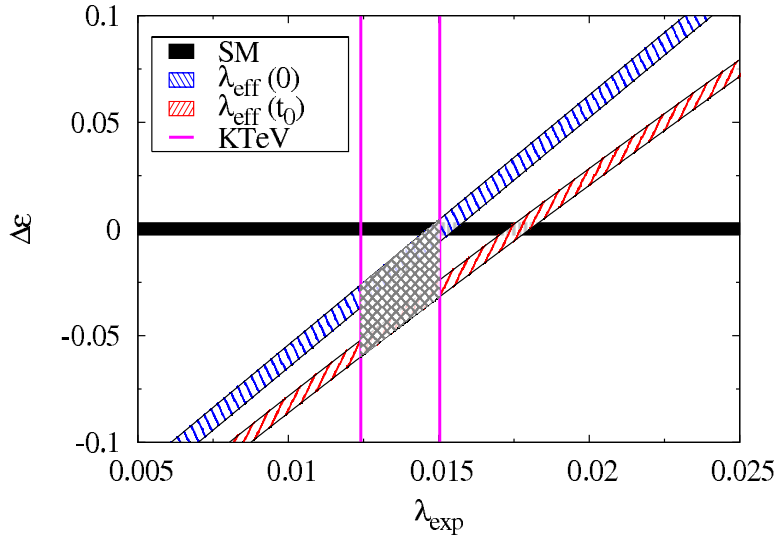}
\caption{Impact of KTeV data on RHCs. Horizontal line: SM, $\Delta \epsilon = \tilde{\Delta}_{CT} = \pm~0.0028$,
vertical lines: KTeV measurements of $\lambda_0$. Top stripped curve: $\lambda_{exp}=\lambda_{eff}(0)$  and bottom
stripped curve: $\lambda_{exp}=\lambda_{eff}(t_0)$ with uncertainties from branching ratios and from G(t) added in quadrature.}    
\label{RHCs}
\end{center}
\end{figure}

      {\bf V. }   In existing fixed target experiments, one usually does not know the
energy of the decaying $K_{L}$ and for this reason it is difficult to
reconstruct the $t = (p_K - p_{\pi})^2$  distribution\footnote {In principle, this difficulty
should not exist in the KLOE experiment.}. The parametrization
of  the two form factors then becomes of prime importance. On the other
hand, it is
difficult to justify a given parametrization experimentally otherwise than
ad hoc,
through a $\chi^2$ of a global fit of measured decay distributions. Under
these circumstances the interpretation of the measured parameters may become
somewhat ambiguous. Keeping in mind that in actual experiments it may be
difficult to determine more than one parameter in the scalar form factor
$f(t)$~\cite{Alexopoulos:2004sy}, either the linear parametrization
\begin{equation}
f(t) = 1 + \lambda \frac{t}{m_{\pi}^2}~,
\label{linear parametrization}
\end{equation}
or the ''pole parametrization''
\begin{equation}
f(t) = \frac{M^2_S}{M^2_S -t}~,
\label{pole parametrization}
\end{equation}
is being standardly used, with a value for $\lambda$ and/or $M_S$ as
the outcome of the fit. To the extent that the formulae
Eq.~($\ref{linear parametrization}$) or~($\ref{pole
parametrization}$) do not correctly account for the curvature of the
form factor in the physical region, it may be questionable whether the
parameters $\lambda$ or $M_S$ measured in this way can indeed be
interpreted as $ m_{\pi}^2 f'(0)$ or as the position of a pole,
respectively.\newline  
\indent One of the obvious advantages of the dispersive
representation Eq.~($\ref{Dispf}$) is that it describes both the linear
slope $\lambda$ and the curvature in terms of the single parameter
$\mathrm{ln C}$. Consider the Taylor expansion
\begin{equation}
f(t) = 1 + \lambda \frac{t}{m_{\pi}^2} + \frac{1}{2} \lambda'
(\frac{t}{m_{\pi}^2})^2  + \ldots~.
\label{taylor}
\end{equation}
The linear slope is
\begin{eqnarray}
\lambda = \frac{m_{\pi}^2}{\Delta_{K \pi}}(\mathrm {ln C} - d )~,
\end{eqnarray}
whereas the curvature reads
\begin{eqnarray}
\lambda' = \lambda^2  - 2 \frac{m_{\pi}^4}{\Delta_{K\pi}} G'(0) = \lambda^2  + (4.16 \pm 0.50)\times 10^{-4} 
\label{curvature}
\end{eqnarray}
It is worth stressing that the lower bound $ \lambda' > \lambda^2 $ is
a general consequence of the positive sign of the phase
$\delta_{K\pi}(s)$ at low energies. Furthermore, even the pole
parametrization Eq.~($\ref{pole parametrization}$) for which $\lambda'
=2 \lambda^2$ underestimates the curvature Eq.~(\ref{curvature}) given by
the dispersive theory, unless $\lambda > 0.020$ or $M_S < 1$~GeV (which
seems excluded by the KTeV results~\cite{Alexopoulos:2004sy}). Note that truncating
the Taylor expansion Eq.~(\ref{taylor}) at the quadratic order and using 
Eq.~(\ref{curvature}) for the curvature, represents by itself an excellent 
approximation in the physical region though not as good as Eq.~(\ref{GP}).\\
\indent Let us illustrate and quantify the difficulty one encounters when the
KTeV result $\lambda_{exp} = 0.01372 \pm 
0.00131$~\cite{Alexopoulos:2004sy} based on the linear parametrization
of $f(t)$ and a quadratic parametrization of $f_{+}(t)$ is converted
into an information on the RHCs. For this purpose we define the
effective (t-dependent) slope
\begin{equation}
 f(t) = 1 + \lambda_{eff}(t) \frac{t}{m_{\pi}^2}.
\end{equation}
\noindent Since $f(t)$ is convex, $\lambda_{eff}(t)$ grows as $t$ increases from
$0$ to $t_0$ . For every fixed $ 0 \le t \le t_0$, $\lambda_{eff}$ is
a function of $\mathrm{ln C}$ or, using Eq.~(\ref{lnC}), of the variable $\Delta
\epsilon = 2 (\epsilon_S - \epsilon_{NS}) + \tilde \Delta_{CT}$ which
we want to constrain. For the extreme cases $t = 0$ and $t = t_0$,
these two curves are displayed in Fig.~\ref{RHCs} together with the
corresponding uncertainties. The problem is that nothing in the KTeV
analysis tells whether the measured value $\lambda_{exp}$ should be
interpreted as $\lambda_{eff}(0)$, $\lambda_{eff}(t_0)$ or
$\lambda_{eff}$ at any other point of the physical region. 
Following whether $\lambda_{exp}$ is identified either with the slope at
$t=0$ or with the slope at $t=t_0$, one obtains respectively:
\begin{equation}
\lambda_{exp}=\lambda_{eff}(0) \Rightarrow \mathrm{lnC}=0.2005 \pm 0.0153 \pm 0.0040~,
\label{lambdaeff}
\end{equation}
\begin{equation}
\lambda_{exp}=\lambda_{eff}(t_0) \Rightarrow \mathrm{ln C} = 0.1748 \pm 0.0141 \pm 0.0019~,
\label{lambda0}
\end{equation}
In both cases, the first error stems from the KTeV error bars whereas the second quoted
error originates from $\delta \mathrm{G(t)}$. The difference of these two
extreme values of $\mathrm{ln C}$ reflects the ambiguity arising because the
theoretically inappropriate linear parametrization has been used in
the analysis of KTeV data. It has little to do with a genuine experimental
or theoretical error. Awaiting an unambiguous measurement of $\mathrm{lnC}$
based on a faithfull parametrization of $f(t)$, we use an admittedly arbitrary
definition of the central value of $\mathrm{lnC}$ identifying $\lambda_{exp}$ with the average 
effective slope $\bar{\lambda}_{eff}$:
\begin{equation}
\lambda_{exp} = \bar \lambda_{eff} = \frac{1}{t_0}\int_{0}^{t_0}dt
\,\lambda_{eff} (t)~.
\label{lambdamoy}
\end{equation}
This gives
\begin{equation}
\mathrm{ln C} = 0.188 \pm 0.015 \pm 0.003 \pm 0.013~,
\label{lnCfinal}
\end{equation}
where to the experimental error and theoretical error arising from
$\delta \mathrm{G}$ we have added the uncertainty due to the inadequate
parametrization of the form factor used in the analysis (the half of the difference 
between Eq.~(\ref{lambdaeff}) and Eq.~(\ref{lambda0})). This last "parametrization
uncertainty" is not a gaussian error and it should not be added in quadrature.
Yet it is almost as big as the genuine experimental error. The corresponding constraint 
on the parameter $\Delta \epsilon$ and on the RHCs is obtained comparing Eq.~($\ref{lnCfinal}$)
with Eq.~($\ref{lnC}$):
\begin{equation}     
\Delta \epsilon = 2(\epsilon_S - \epsilon_{NS}) + \tilde \Delta_{CT} = - 0.029 \pm 0.015
\pm 0.003 \pm 0.013~.
\end{equation}

\noindent Thanks to the "parametrization uncertainty" this result for $\Delta \epsilon$ is still compatible with zero, i.e. with
the SM. (Recall that $\tilde{\Delta}_{CT}$ is expected of the order $\sim 2.8~10^{-3}$~\cite{Gasser:1984ux}). 
This rough result $\Delta \epsilon = -0.03 \pm 0.03$ summarized in Fig.~\ref{RHCs} is rather robust.
In particular, it
does not depend on the detailed prescription Eq.~($\ref{lambdamoy}$) of defining the 
central value of $\mathrm{lnC}$.
                                
                        {\bf VI} We finally come to the charged K-decay mode $K^+ \to \pi^0\mu\nu$. In full analogy to the neutral Kaon mode
(cf Eq.~\ref{C}), we can define the corresponding Callan-Treiman discrepancy
$\Delta^{K^+}_{CT}$. Using the E865 result~\cite{Sher:2003fb} 
for $|f_+^{K^+\pi^0}(0) \Veff_{eff}^{us}|$, we obtain:
\begin{equation}
\mathrm{lnC}^{K^+}=0.1798 \pm 0.0105 + \tilde\Delta_{CT}^{K^+} + 2 (\epsilon_S - \epsilon_{NS})+\calo(\epsilon^2)~, 
\label{lnCK+}
\end{equation}
where $\tilde\Delta_{CT}^{K^+}= \Delta_{CT}^{K^+}/B_{exp}^{K^+}= \Delta_{CT}^{K^+}/1.20$. 
Note that the experimental uncertainty is about 3 times larger than in the neutral case, Eq.~(\ref{lnC}), and of the same order of magnitude
as the expected effect of RHCs.
Furthermore, the effect of isospin breaking on $\Delta_{CT}^{K^+}$ due to $m_d-m_u$ is amplified by small denominators 
arising from $\pi^0$-$\eta$ mixing.  
Evaluating $\Delta_{CT}^{K^+}$ at order $\calo(p^4,(m_d-m_u)p^2,e^2p^2)$ within $\chi PT$ using the results of~\cite{Cirigliano:2001mk}
we find an increase of a few percent. (A similar increase is already observed in the decay rate~\cite{Cirigliano:2004pv}.)
Varying the parameters as described for $K^0$, one gets:
$0.02<\Delta_{CT}^{K^+}<0.05$.
Thus compared with the neutral case Eq.~(\ref{lnC}), the decrease of the first number on the RHS of Eq.~(\ref{lnCK+})
(reflecting the decrease of $B_{exp}^{K^+}$) could be compensated by a larger value of $\Delta_{CT}^{K^+}$.
It thus seems that the analysis of RHCs from the $K^+_{\mu3}$ experiment~\cite{Yushchenko:2003xz} 
is more involved requiring among other things a better knowledge of the isospin breaking parameter $\epsilon^{(2)}= \frac{\sqrt 3}{4} \frac {1}{R}$. 

             {\bf VII.} a) In this paper, we have constructed an accurate low-energy dispersive representation of the scalar $K\pi$ form factor in terms 
of a single parameter $C=f(\Delta_{K\pi})$ which describes both its slope and its curvature. The result, in a form ready to be
used in a $K^L_{\mu3}$ decay analysis is presented in Eqs.~($\ref{Dispf}$), ($\ref{GP}$) and in Table~1. Alternatively, an even simpler
but somewhat less accurate quadratic parametrization could be used provided the slope and curvature are related by 
Eq.~($\ref{curvature}$).\\
b) We have shown that a measurement of lnC at the 0.01 level would provide a significant test of direct electroweak couplings of
right-handed quarks to the standard $W$-boson. So far no such a test is available beyond the specific framework of left-right
symmetric extensions of the Standard Model.\\
c) Using the recent KTeV data~\cite{Alexopoulos:2004sy} we have shown that the linear parametrization of the scalar form factor leads to 
a "parametrization uncertainty" in lnC comparable with the actual experimental error $\pm 0.015$. This loss of information can be avoided
taking into account the curvature of the form factor as suggested in this paper.\\
d) In parallel, matching the measured parameter $C=f(\Delta_{K \pi})$, the slope and the curvature of the scalar form factor
with the two-loop $\chi PT$~\cite{Bijnens:2003uy} would help the  assessment of the CKM element $V_L^{us}$.\\   
\\
\noindent\underline{{\bf Acknowledgements:}}
\noindent
We thank A.~Ceccucci, H.~Leutwyler, B.~Moussallam and J.A.~Oller for 
their interest, suggestions
and help. Work supported in part by the EU RTN 
contract HPRN-CT-2002-00311 (EURIDICE) and by the EU I3HP contract 
RII3-CT-2004-506078.


\end{document}